\documentclass[sigconf]{acmart}

\AtBeginDocument{%
  }

\setcopyright{acmlicensed}
\copyrightyear{2018}
\acmYear{2018}
\acmDOI{XXXXXXX.XXXXXXX}

\acmConference[Conference acronym 'XX]{Make sure to enter the correct
  conference title from your rights confirmation email}{June 03--05,
  2018}{Woodstock, NY}

\acmISBN{978-1-4503-XXXX-X/2018/06}

\usepackage{tabularx}
\usepackage{multirow}
\usepackage[table]{xcolor}
\definecolor{oursblue}{RGB}{220,235,255}
\definecolor{lightgrayrow}{RGB}{245,245,245}

\usepackage{tabularx}
\usepackage[table,xcdraw]{xcolor} 

\begin{document}

\title{OneModel: A Unified Foundation for Platform-Scale Multi-Scenario Ranking}
\renewcommand{\shorttitle}{OneModel}

\author{Yinqi Zhang}
\affiliation{%
  \institution{Algorithm, Xiaohongshu}
  \city{Shanghai}
  \country{China}
}

\author{Peiyu Hu}
\authornote{Work as an intern at Xiaohongshu.}
\affiliation{%
  \institution{Algorithm, Xiaohongshu}
  \city{Shanghai}
  \country{China}
}

\author{Yuntian Tang}
\authornotemark[1]
\affiliation{%
  \institution{Algorithm, Xiaohongshu}
  \city{Shanghai}
  \country{China}
}

\author{Siying Gu}
\authornotemark[1]
\affiliation{%
  \institution{Algorithm, Xiaohongshu}
  \city{Shanghai}
  \country{China}
}

\author{Jiahao Liang}
\authornotemark[1]
\affiliation{%
  \institution{Algorithm, Xiaohongshu}
  \city{Shanghai}
  \country{China}
}

\author{Longxin Kou}
\authornotemark[1]
\affiliation{%
  \institution{Algorithm, Xiaohongshu}
  \city{Shanghai}
  \country{China}
}

\author{Haiqing Hu}
\authornotemark[1]
\affiliation{%
  \institution{Algorithm, Xiaohongshu}
  \city{Shanghai}
  \country{China}
}

\author{Shuman Zhuang}
\affiliation{%
  \institution{Inference Infra, Xiaohongshu}
  \city{Beijing}
  \country{China}
}

\author{Yubin Xu}
\affiliation{%
  \institution{Inference Infra, Xiaohongshu}
  \city{Beijing}
  \country{China}
}

\author{Chenggen Sun}
\affiliation{%
  \institution{Inference Infra, Xiaohongshu}
  \city{Beijing}
  \country{China}
}

\author{Bin Ye}
\affiliation{%
  \institution{Training Infra, Xiaohongshu}
  \city{Shanghai}
  \country{China}
}

\author{Donghui Xu}
\affiliation{%
  \institution{Training Infra, Xiaohongshu}
  \city{Shanghai}
  \country{China}
}

\author{Zhaoyu Liu}
\affiliation{%
  \institution{Training Infra, Xiaohongshu}
  \city{Shanghai}
  \country{China}
}

\author{Jiang Rong}
\affiliation{%
  \institution{Algorithm, Xiaohongshu}
  \city{Beijing}
  \country{China}
}

\author{Yuting Jia}
\affiliation{%
  \institution{Training Infra, Xiaohongshu}
  \city{Shanghai}
  \country{China}
}

\author{Zhaokai Luo}
\affiliation{%
  \institution{Inference Infra, Xiaohongshu}
  \city{Beijing}
  \country{China}
}

\author{Leilei Ma}
\affiliation{%
  \institution{Inference Infra, Xiaohongshu}
  \city{Beijing}
  \country{China}
}

\author{Yiying Xie}
\authornote{Corresponding Author.}
\email{yiyingxie@xiaohongshu.com}
\affiliation{%
  \institution{Algorithm, Xiaohongshu}
  \city{Shanghai}
  \country{China}
}

\author{Yao Hu}
\authornote{Team Leader.}
\affiliation{%
  \institution{Xiaohongshu}
  \city{Beijing}
  \country{China}
}

\renewcommand{\shortauthors}{Zhang et al.}

\begin{abstract}

Platform-scale recommender systems often span multiple business streams such as organic recommendation, advertising, and merchant services, where user behaviors form a continuous cross-stream trajectory. Maintaining separate ranking systems fragments user representations and increases engineering cost. We propose \textbf{OneModel}, a unified framework for multi-stream final ranking. OneModel maps heterogeneous behaviors into shared event sequences, learns long-context user representations with an action-oriented backbone, and introduces \emph{Scenario-aware Information Modulation} to balance cross-stream transfer and stream-specific specialization. For production deployment, OneModel further adopts stratified user representation, multi-objective training, and optimized online serving with feature decomposition, user feature prefetching, shared user-tower computation, and graph-level inference optimization. We deploy OneModel in production at \emph{Xiaohongshu}, where it delivers consistent offline gains over strong baselines and scales favorably with context length and model capacity. Online A/B tests improve Time Spent by \textbf{+0.33\%} and Engagement by \textbf{+1.25\%} in Explore Feed, lift advertising value by \textbf{+3.43\%} and CTR by \textbf{+8.18\%} in Feed Advertising, and raise DGMV by \textbf{+1.1867\%} and GPM by \textbf{+2.1585\%} in Merchant Recommendation, validating unified multi-stream ranking as an effective production foundation.

\end{abstract}



\received{20 February 2007}
\received[revised]{12 March 2009}
\received[accepted]{5 June 2009}

\maketitle

\section{Introduction}

Recommender systems form the core infrastructure of modern Internet platforms and directly affect content distribution, commercial conversion, and user experience~\cite{covington2016youtube,cheng2016wide,guo2017deepfm,zhou2018deep, rec1,rec2,rec3, hu2026modular}. Traditional industrial ranking followed a discriminative paradigm with increasingly sophisticated feature interactions~\cite{rendle2010fm,juan2016ffm,cheng2016wide,guo2017deepfm,lian2018xdeepfm,wang2017dcn}. More recent generative recommendation and large-scale sequence ranking systems such as HSTU and GenRank~\cite{hstu,genrank,mtgr,liu2024kuaiformer,survey1,survey2,survey3} shift the focus to long action-aware behavior sequences and reveal a clearer scaling path that benefits from larger data, longer contexts, and higher-capacity models.

This scaling path relies on a set of heavy capabilities, including long-history memory, stronger content understanding, large embeddings, and multi-modal modeling~\cite{survey3,survey4,rec2,scaling1}, which are increasingly necessary for pushing the performance ceiling beyond shallow feature crossing. These capabilities, however, are also expensive to train, serve, and iterate, since they consume substantial compute, storage, and engineering resources.

\begin{figure}[t]
    \centering
    \includegraphics[width=0.95\linewidth]{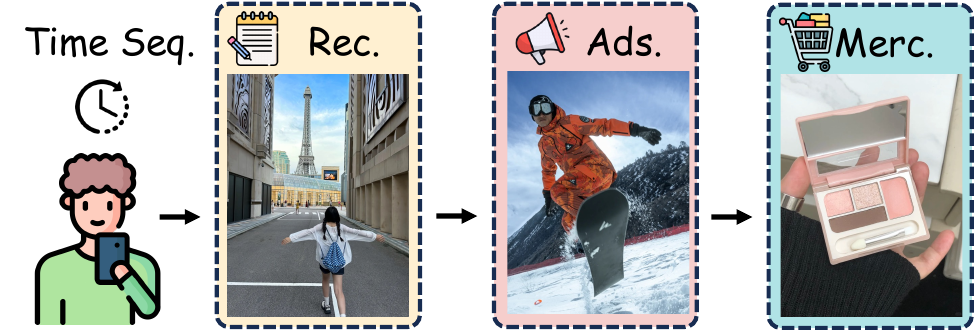}
    \caption{A typical user's sequential behavior log across multiple scenarios in Xiaohongshu.}

    \label{fig:intro}
\end{figure}

It is therefore increasingly important to reuse these heavy capabilities across businesses rather than rebuild them separately. In compound platforms such as Xiaohongshu, users naturally interact with three business streams, namely organic recommendation, advertising, and merchant services, within one continuous journey (Figure~\ref{fig:intro}); we refer to these collectively as multi-stream behaviors. These streams share users, contents, and intent signals: content consumption often reveals purchase intent, ad responses reflect commercial preference, and merchant behaviors provide high-value supervision for general interest modeling. A unified multi-stream model can both reuse such cross-stream information for better performance and replace multiple isolated long-sequence models, embedding systems, and content-understanding pipelines with a single shared foundation.

Unifying multiple business streams, however, is non-trivial. The streams use different feature schemas and value signals such as interest matching for recommendation, ROI for advertising, and conversion and GMV for merchant; naive parameter sharing risks cross-business interference among correlated but conflicting objectives; the unified model still needs to support business-specific specialization without degenerating into fully separated towers; and the shared foundation must satisfy strict industrial serving constraints under which long-context modeling cannot be recomputed for every request.

To address these challenges, we propose \textbf{OneModel}, a unified generative ranking framework for platform-scale multi-stream ranking. To handle heterogeneous inputs, OneModel maps multi-stream behaviors into a shared event sequence through scenario-aware feature projection and structural context encoding. To reuse scalable heavy capabilities, OneModel builds an action-oriented long-context sequence backbone that learns shared user preference and content semantics from mixed histories. To mitigate cross-stream interference, OneModel introduces \emph{Scenario-aware Information Modulation (SAIM)} and modular multi-objective optimization, so that shared representations adapt to different business goals. To support production deployment, OneModel further adopts stratified user representation, incremental user-state caching, and vectorized candidate scoring to cut online cost. We deploy OneModel in the production system of Xiaohongshu, where it delivers consistent online gains across the three business streams, including improvements in engagement, advertising value, and merchant-side conversion efficiency.

The main contributions of this paper are summarized as follows:

\begin{itemize}
\item We propose \textbf{OneModel}, a unified platform-scale multi-stream ranking framework that consolidates reusable user and content understanding into a shared foundation across organic recommendation, advertising, and merchant services, and thereby improves information reuse and resource utilization.

\item We design unified cross-stream representation, an action-oriented long-context backbone, \emph{Scenario-aware Information Modulation (SAIM)}, stratified user modeling, modular multi-objective optimization, and decoupled online serving, which together balance sharing, specialization, and deployment cost.

\item We deploy OneModel in real-world production systems and run extensive offline and online A/B tests, which confirm its gains in recommendation engagement, advertising value, and merchant-side conversion under serving constraints.
\end{itemize}

\section{Related Work}

\textbf{Long-context and generative ranking.}
Industrial ranking evolved from discriminative utility estimation with feature interaction learning and CTR prediction~\cite{rendle2010fm,juan2016ffm,cheng2016wide,guo2017deepfm,lian2018xdeepfm,wang2017dcn,zhou2018deep,huang2013learning} to sequential recommendation that models ordered user histories~\cite{GRU4Rec,SASRec,Bert4rec,survey1,survey4}. Earlier work used user behavior sequence modeling for target-aware interest modeling, interest evolution, and long-term preference learning~\cite{din,dien,mimn,sim}. More recent Transformer-based and generative ranking systems, such as HSTU, GenRank, MTGR, and related industrial frameworks, further scaled this direction with action-aware or event-based long-context modeling~\cite{hstu,genrank,mtgr,liu2024kuaiformer, hu2026hierarchical, liu2026fedcgr}, while other large ranking models combined feature interaction, sequence modeling, and model scaling in stronger backbones~\cite{rankmixer,zhang2026onetrans,jiang2026tokenmixer}. Most of these models, however, still targeted a single scenario or homogeneous objectives.

\textbf{Multi-task and cross-scenario recommendation.}
Multi-task and cross-scenario learning methods shared knowledge across related tasks or business scenarios through shared-bottom structures, expert routing, scenario-specific towers, or adaptive sharing mechanisms~\cite{caruana1997multitask,tang2020progressive,wang2022causalint,liu2024multi,song2024multiscenario,yi2025adaptive,xu2025cross}. To mitigate the negative transfer caused by heterogeneous objectives and distributions, recent studies also investigated how to control shared and private information~\cite{bai2022contrastive,zhou2023fdn} and how to exploit shared user interests across service contexts~\cite{xu2025cross}. These methods, however, often relied on shallow shared representations, separated scenario-specific modules, or offline user modeling, which makes them less suited for modeling long mixed multi-stream histories as one continuous trajectory for production final ranking.

\textbf{Industrial unified ranking systems.}
Recent industrial generative recommendation frameworks explored scalable sequence modeling, multi-objective ranking, and multi-business prediction~\cite{sortgen,pantheon,mbgr,tencentgr}, reflecting a broader trend toward large-scale sequence-based and unified recommendation systems~\cite{onepiece,zhang2026onetrans,yan2023hyformer}. Most of them, however, focused on semantic-ID generation, list-level re-ranking, multi-objective sorting, or single advertising scenarios. In contrast, OneModel targets platform-scale final ranking over long mixed multi-stream histories, and uses a shared long-context backbone with scenario-aware modulation to balance transfer and specialization under serving constraints.

\section{Preliminary}
\label{sec:preliminary}

We formulate platform-scale multi-stream ranking over users $\mathcal{U}$, items $\mathcal{I}$, and business streams $\mathcal{S}=\{\text{recommendation}, \text{advertising}, \\ \text{merchant}\}$.
Each item $i\in\mathcal{I}$ carries a stream origin $s_i\in\mathcal{S}$ and heterogeneous features $\mathbf{x}_i$.
For user $u$, we represent the cross-stream behavior history as a time-ordered sequence
$\mathcal{H}_u=\{e_1,e_2,\dots,e_L\}$, where each event
$e_j=(i_j,s_j,a_j,t_j)$ records the interacted item, stream, action type, and timestamp, with $t_1<t_2<\cdots<t_L$.

OneModel learns a unified function $f_\theta$ that maps $\mathcal{H}_u$ to a user representation $\mathbf{u}$ for downstream prediction, and supports two correlated objectives.
The first objective is generative retrieval, which predicts the next interacted item from the unified item pool by modeling $P(i_{L+1}\mid\mathcal{H}_u)$.
The second objective is stream-specific ranking, which estimates the target action probability for a candidate item $i$ under stream $s$ as $\hat{y}=g(\mathbf{u},\mathbf{x}_i,s)$.
Such a setting requires the unified model to handle heterogeneous feature spaces, reduce cross-stream negative transfer, and satisfy real-time serving constraints.

\section{Methodology}
\label{sec:method}


\subsection{Unified Cross-scenario Representation Learning}
\label{sec:representation_learning}

\begin{figure}[t]
    \centering
    \includegraphics[width=0.85\linewidth]{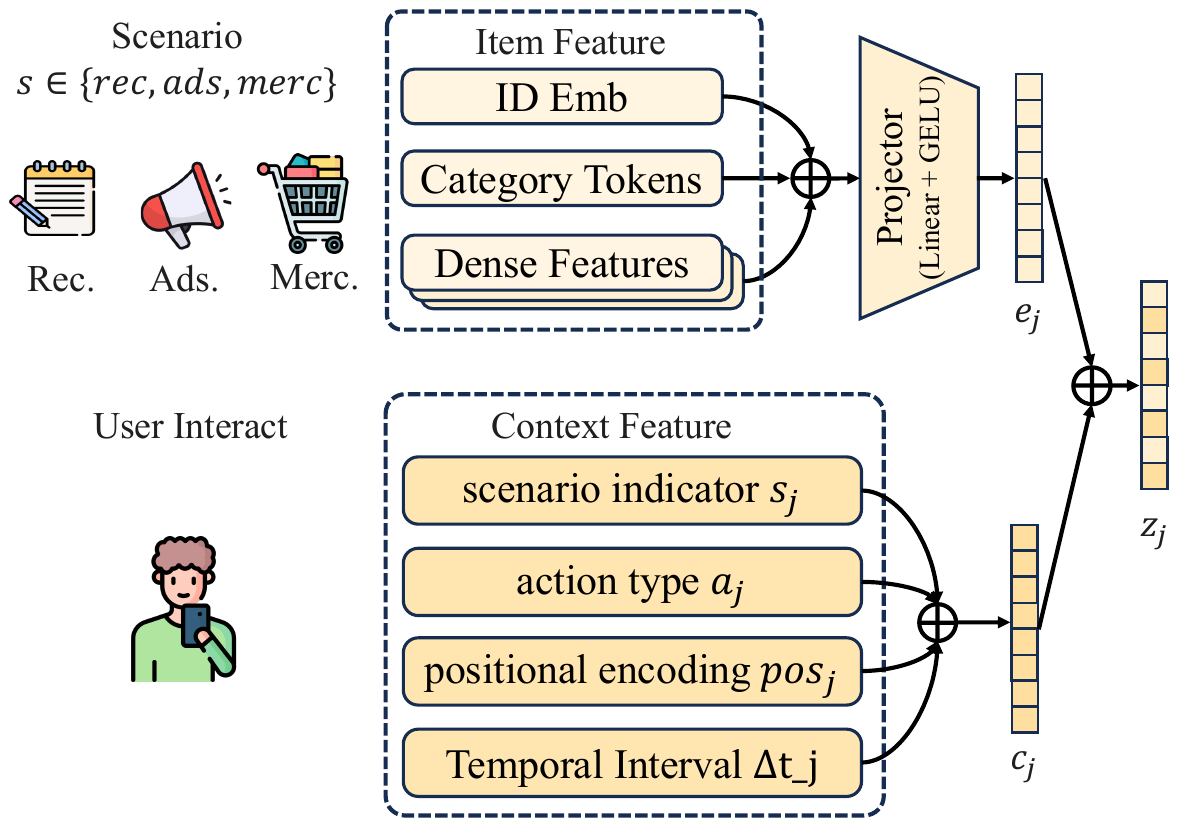}
    \caption{Unified cross-scenario representation learning via scenario-aware projection and structural context integration.}
    \label{fig:data}
\end{figure}

To unify heterogeneous inputs, OneModel maps scenario-specific item features and behavioral contexts into a shared event-token space. As shown in Figure~\ref{fig:data}, this module contains two steps: scenario-aware feature projection and structural context integration.

\subsubsection{Scenario-aware Feature Projection}

Given an item $i$ from business stream $s\in\mathcal{S}$, we concatenate its heterogeneous raw features, including ID embeddings, category tokens, and normalized dense features, into $\mathbf{x}_i^{(s)}$. To preserve stream-specific feature structures while mapping them into a unified space, we apply a lightweight scenario-specific projection:
\begin{equation}
    \mathbf{e}_i = \phi\left(\mathbf{W}^{(s)}\mathbf{x}_i^{(s)} + \mathbf{b}^{(s)}\right),
    \label{eq:projection}
\end{equation}
where $\mathbf{W}^{(s)}$ and $\mathbf{b}^{(s)}$ are scenario-specific parameters, and $\phi(\cdot)$ is the activation function. The resulting $\mathbf{e}_i$ provides a comparable item representation across scenarios.

\subsubsection{Structural Context Integration}

To encode each user interaction as a structured event token, we inject scenario, action, temporal, and positional context into the item representation:
\begin{equation}
    \mathbf{z}_j = \mathbf{e}_{i_j} + \underbrace{\text{Embed}(s_j, a_j, \Delta t_j, pos_j)}_{\text{Context Embedding } \mathbf{c}_j},
    \label{eq:structural_token}
\end{equation}
where $s_j$ is the scenario indicator, $a_j$ is the action type, $pos_j$ is the position, and $\Delta t_j=t_j-t_{j-1}$ is the time interval.
We encode $\Delta t_j$ with multi-frequency sinusoidal embeddings and use learnable embeddings for discrete attributes such as scenario and action type.
These structured tokens provide the long-sequence backbone with comparable cross-scenario inputs while preserving behavioral semantics such as recency, action intent, and scenario identity.

\subsection{Hierarchical Cross-scenario Sequence Modeling}
\label{sec:sequence_modeling}

\begin{figure}[t]
    \centering
    \includegraphics[width=0.8\linewidth]{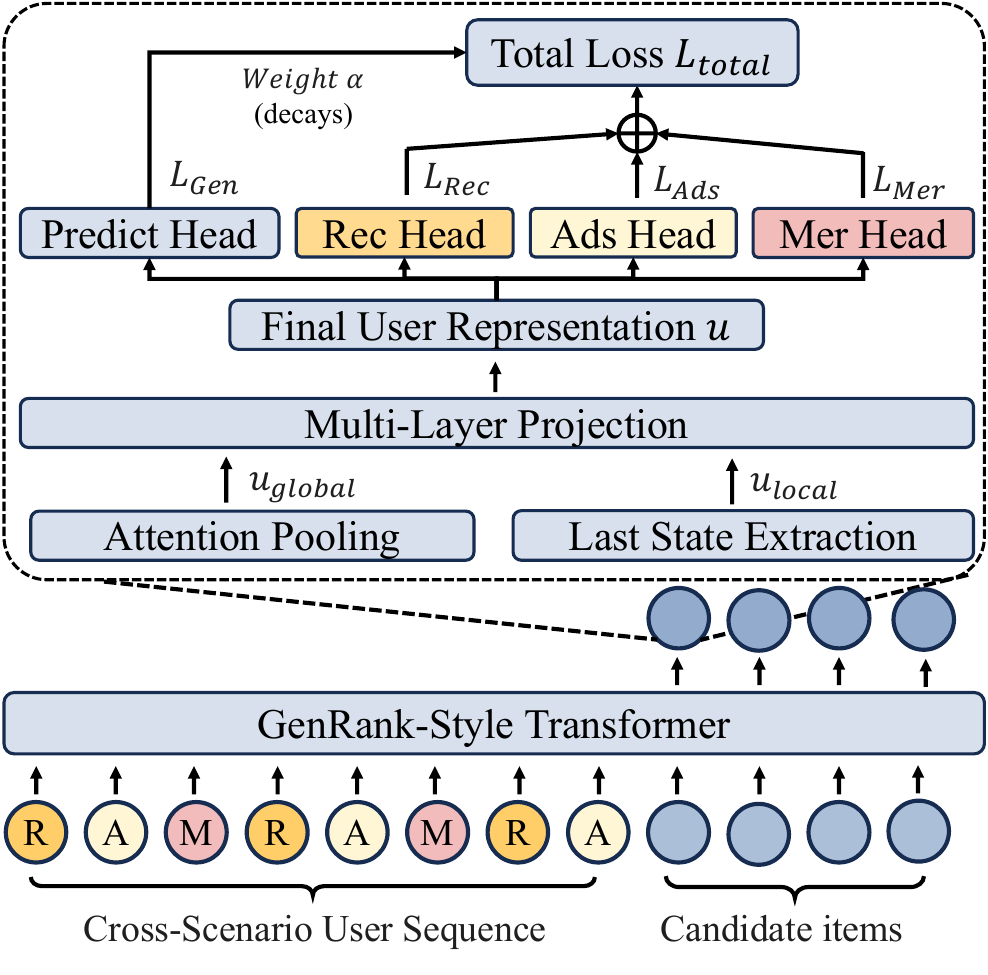}
    \caption{Overall architecture of hierarchical cross-scenario sequence modeling. OneModel uses explicit item-context interaction and an action-oriented decoder with scenario-aware modulation and stratified user representation, followed by multi-task heads for the three business streams.}
    \label{fig:arc}
\end{figure}

To model long multi-stream histories efficiently, OneModel builds a unified sequence module on a GenRank-style action-oriented backbone, as shown in Figure~\ref{fig:arc}. The module contains four designs: action-oriented tokenization for reducing effective sequence length, explicit item-context interaction before sequence encoding, a unified causal decoder with candidate masking, scenario-aware information modulation, and stratified user representation.

\textbf{Action-oriented organization.}
Following GenRank, OneModel treats actions as prediction targets and items as contextual signals, avoiding the doubled length caused by explicitly interleaving item and action tokens. Compared with HSTU-style tokenization, which preserves separate item and action tokens and can retain fine-grained interaction semantics, this organization substantially shortens the effective sequence length and improves serving efficiency. Different from directly compressing item-context features into sequence tokens, OneModel first performs explicit item-context interaction before sequence encoding, which strengthens candidate-side feature crossing while keeping the sequence backbone compact.
Given historical interactions $\{(x_i,a_i,t_i)\}_{i=1}^{N}$, we model
\begin{equation}
p(a_k \mid x_1,a_1,\ldots,x_k).
\label{eq:action_dist}
\end{equation}
Each historical token is represented as
\begin{equation}
\mathbf{e}_i=\phi(x_i)+\varphi(a_i)+\mathbf{c}_i,
\label{eq:hist_token}
\end{equation}
where $\phi(\cdot)$ is the unified item representation, $\varphi(\cdot)$ is the action embedding, and $\mathbf{c}_i$ contains structural context.
For candidate items $\{x_j\}$, we use a masked action embedding $\mathbf{m}$ to prevent target-action leakage:
\begin{equation}
\mathbf{e}^{\text{cand}}_j=\phi(x_j)+\mathbf{m}+\mathbf{c}^{\text{cand}}_j .
\label{eq:cand_token}
\end{equation}

\textbf{Unified cross-scenario decoder.}
We feed the token sequence $\mathcal{E}=[\mathbf{e}_1,\ldots,\mathbf{e}_N,\mathbf{e}^{\text{cand}}_1,\ldots,\mathbf{e}^{\text{cand}}_M]$ into a shared causal Transformer decoder, and apply both causal masking and candidate masking, where the former preserves temporal order and the latter blocks attention among candidates in the same request:
\begin{align}
\mathbf{X}^{(l)} &= \mathbf{H}^{(l-1)} + \text{Attn}\!\left(\text{Norm}(\mathbf{H}^{(l-1)}); \mathbf{M}_{\text{causal}}, \mathbf{M}_{\text{cand}}\right), \\
\mathbf{H}^{(l)} &= \mathbf{X}^{(l)} + \text{FFN}\!\left(\text{Norm}(\mathbf{X}^{(l)})\right).
\label{eq:genrank_block}
\end{align}
For efficient position and time encoding, we define the structural context as
\begin{equation}
\mathbf{c}_i=\mathbf{E}^{pe}_i+\mathbf{E}^{ri}_i+\mathbf{E}^{rt}_i+\text{Embed}(s_i),
\label{eq:pos_time_ctx}
\end{equation}
where $\mathbf{E}^{pe}_i$, $\mathbf{E}^{ri}_i$, and $\mathbf{E}^{rt}_i$ encode interaction position, request index, and bucketed pre-request time gap.
We additionally use an ALiBi-style linear bias to provide relative temporal inductive bias without quadratic bias-table overhead.

\begin{figure}[t]
    \centering
    \includegraphics[width=0.95\linewidth]{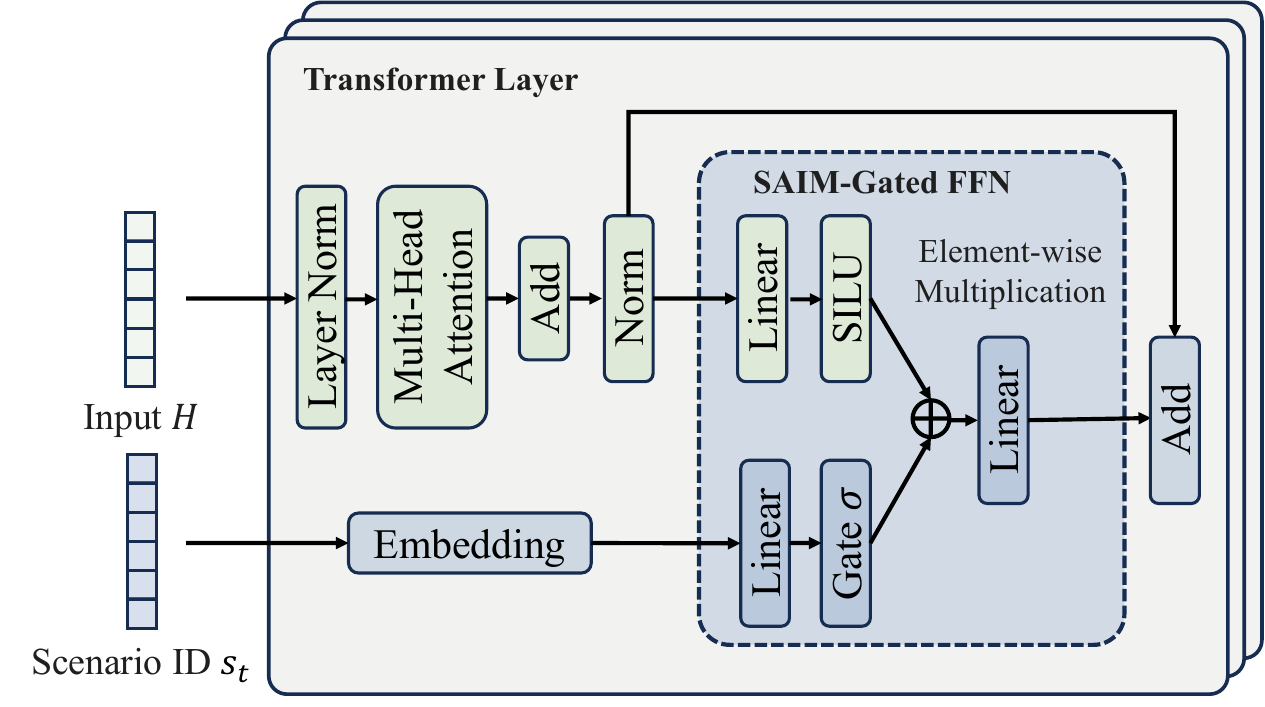}
    \caption{Scenario-aware Information Modulation (SAIM).}
    \label{fig:ab}
\end{figure}

\textbf{Scenario-aware information modulation.}
To reduce interference among mixed-stream tokens, we insert a lightweight scenario-conditioned gate into the FFN, as shown in Figure~\ref{fig:ab}.
For token $t$ with scenario $s_t$, the gate is
\begin{equation}
\mathbf{g}_t=\sigma(\mathbf{W}_g\mathbf{e}_{s_t}+\mathbf{b}_g),
\qquad \mathbf{g}_t\in(0,1)^{d_f},
\label{eq:SAIM_gate}
\end{equation}
and the FFN output becomes
\begin{equation}
\text{FFN}^{(s_t)}(\mathbf{h}_t)=
\left(\text{SiLU}(\mathbf{W}_1\mathbf{h}_t)\odot\mathbf{g}_t\right)\mathbf{W}_2+\mathbf{b}_2 .
\label{eq:SAIM_ffn}
\end{equation}
This design keeps one shared computation graph while allowing stream-dependent channel modulation.

\textbf{Stratified user representation.}
To capture both long-term preference and short-term intent, we combine global attention pooling with the final local state:
\begin{equation}
    \mathbf{u}_{global}=\sum_{t=1}^{L}\alpha_t\mathbf{h}^{(L)}_t,\quad
    \alpha_t=\text{Softmax}\!\left(\mathbf{v}^{\top}\tanh(\mathbf{W}_a\mathbf{h}^{(L)}_t)\right),
    \label{eq:global_pool}
\end{equation}
\begin{equation}
    \mathbf{u}=\text{MLP}\left([\mathbf{u}_{global}\parallel\mathbf{u}_{local}]\right),
    \label{eq:user_fusion}
\end{equation}
where $\mathbf{u}_{local}=\mathbf{h}^{(L)}_L$.
Downstream multi-task prediction heads then consume the fused representation $\mathbf{u}$.
\subsection{End-to-End Multi-scenario Optimization}
\label{sec:optimization}

To jointly optimize shared user modeling and stream-specific ranking, OneModel trains all components end-to-end with a unified multi-objective loss that combines self-supervised next-item prediction, supervised stream-specific prediction, and parameter regularization:
\begin{align}
\mathcal{L} &= \alpha \cdot \mathbb{E}_{u} \left[ -\sum_{j=1}^{L-1} \log P(i_{j+1}|\mathcal{S}_{\leq j}) \right] \label{eq:seq_loss} \\
&+ \beta \sum_{k=1}^K \lambda_k \cdot \mathbb{E}_{(u,i)\sim \mathcal{D}_k} \left[ \ell_k(\hat{y}^{(k)}, y^{(k)}) \right] \nonumber \\
&+ \gamma \|\Theta\|_2^2 . \nonumber
\end{align}
The first term learns general sequential preference from mixed histories, the second term aligns the representation with $K$ stream-specific objectives, and the last term regularizes model parameters. To balance representation learning and task alignment, we decay $\alpha$ and grow $\beta$ during training, and set $\lambda_k$ inversely proportional to the validation AUC of stream $k$ so that the model focuses more on harder streams.

To stabilize optimization under heterogeneous objectives, we use cross-stream batch sampling and early-stage gradient isolation. Cross-stream sampling prevents high-frequency streams from dominating the shared backbone, and gradient isolation temporarily detaches backbone gradients from stream-specific towers during warm-up to reduce noisy updates to the shared representations. To further cut ultra-long sequence training cost, we apply selective backpropagation that computes gradients mainly on high-perplexity steps and skips easy positions.

\subsection{Online Serving Strategy}
\label{sec:serving}

To support real-time ranking, OneModel separates long-context user encoding from request-time candidate scoring. The user state $\mathbf{z}_u=f_\theta(\mathcal{H}_u)$ is incrementally updated after new interactions and cached in a low-latency key-value store, avoiding repeated Transformer inference for every request. During ranking, candidate features are stacked as $\mathbf{V}\in\mathbb{R}^{N\times d_v}$ and scored through vectorized computation:
\begin{equation}
\mathbf{s}=G_\Phi(\mathbf{z}_u,\mathbf{V}),
\end{equation}
where $G_\Phi$ denotes the task-specific scoring function. We further apply feature decomposition, user feature prefetching, shared user-tower computation, and graph-level inference optimization to reduce online latency.

\section{Experiments}

We conduct offline and online experiments to answer the following research questions:\\
\textbf{RQ1}: Does OneModel outperform strong single-stream and multi-stream baselines?\\
\textbf{RQ2}: Which training, input, and module designs contribute most?\\
\textbf{RQ3}: How does OneModel scale?\\
\textbf{RQ4}: Is OneModel stable under multi-stream optimization?\\
\textbf{RQ5}: Does OneModel bring online gains in production?\\
\textbf{RQ6}: Can OneModel improve serving efficiency while scaling up?
\subsection{Experimental Setup}

\textbf{Datasets.} We evaluate OneModel on an anonymized 10\% traffic slice from the production platform of \textit{Xiaohongshu}, covering three business streams: organic recommendation, advertising, and merchant services, with a relative traffic ratio of approximately recommendation : advertising : merchant $=$ $1 : 0.2 : 0.15$. We organize user, item, context, and interaction features into timestamp-ordered multi-stream behavior sequences. Due to data confidentiality, we report relative traffic ratios and aggregated evaluation results rather than absolute user, item, or interaction counts.

\noindent\textbf{Metrics.} For offline evaluation, we report AUC and LogLoss on a timestamp-split test set. The main results cover multi-target AUC on Click, Collect, Comment, PageTime, Follow, Hide, and Like. For ablation and scaling analyses, we report AUC or relative AUC changes per experiment. For online evaluation, we report production business metrics and serving efficiency, including user engagement, advertising value, merchant conversion, latency, throughput, and normalized GPU cost.

\noindent\textbf{Implementation.} We implement OneModel in PyTorch and train it with distributed NCCL on TFRecord data. The default configuration uses sequence length $500$, $768$-dimensional item embeddings, $3$ Transformer blocks, $8$ attention heads, FFN width $768$, and dropout $0.2$. We use bf16 mixed precision, flash-attention, gradient checkpointing, and sampled softmax with in-batch negatives. More details are provided in Appendix~\ref{app:implementation_details}.

\subsection{Main Results (RQ1)}
\label{sec:rq1_main_results}

\begin{table*}[!t]
\centering
\small
\setlength{\tabcolsep}{4pt}
\renewcommand{\arraystretch}{1.15}
\caption{Main results under single-stream and multi-stream settings. Rows indicate the business stream, training setting, and model. Columns report AUC for different prediction targets. ``PageTime'' denotes page-time prediction.}
\label{tab:main_results}
\begin{tabularx}{\textwidth}{llXccccccc}
\toprule
\rowcolor{lightgrayrow}
\textbf{Business Stream} & \textbf{Setting} & \textbf{Model}
& \textbf{Click} & \textbf{Collect} & \textbf{Comment} & \textbf{PageTime}
& \textbf{Follow} & \textbf{Hide} & \textbf{Like} \\
\midrule
\multicolumn{10}{l}{\textit{Single-stream setting: advertising-only training and evaluation}} \\
\multirow{3}{*}{Advertising} & \multirow{3}{*}{Single} & HSTU
& 0.7644 & 0.8991 & 0.9006 & 0.6736
& 0.9156 & 0.8782 & 0.8814 \\
& & GenRank
& 0.7648 & 0.8995 & 0.9010 & 0.6740
& 0.9160 & 0.8786 & 0.8818 \\
\rowcolor{oursblue}
& & OneModel
& \textbf{0.7682} & \textbf{0.9032} & \textbf{0.9047} & \textbf{0.6786}
& \textbf{0.9193} & \textbf{0.8822} & \textbf{0.8858} \\
\midrule
\multicolumn{10}{l}{\textit{Multi-stream setting: unified training over all business streams}} \\
Advertising & \multirow{3}{*}{Unified} & \multirow{3}{*}{OneModel}
& 0.7712 & 0.9072 & 0.9082 & 0.6836 & 0.9221 & 0.8864 & 0.8908 \\
Recommendation & &
& 0.7905 & 0.9754 & 0.9609 & 0.7939 & 0.9769 & 0.9353 & 0.9616 \\
Merchant & &
& 0.7819 & 0.9433 & 0.9323 & 0.7261 & 0.9298 & 0.8702 & 0.9358 \\
\bottomrule
\end{tabularx}
\end{table*}

To answer \textbf{RQ1}, we evaluate OneModel under two settings. In the \textit{single-stream} setting, we compare OneModel with HSTU and GenRank on the advertising stream. HSTU uses a long item-action sequence design, while GenRank adopts action-oriented organization as a compact and efficiency-oriented baseline. In the \textit{multi-stream} setting, OneModel is trained on the unified behavior stream and evaluated on per-stream targets. Table~\ref{tab:main_results} reports AUC across seven prediction targets.

On the advertising stream, GenRank slightly outperforms HSTU across all targets, suggesting that action-oriented organization provides a better accuracy--efficiency trade-off under the advertising-only setting. OneModel further surpasses both baselines by combining compact sequence encoding with stronger item-context interaction. On Ads Click AUC, single-stream OneModel improves over GenRank by 3.4\textperthousand, while unified OneModel enlarges the gap to 6.4\textperthousand. Under unified training, OneModel also maintains strong performance across Recommendation, Advertising, and Merchant streams, showing that one shared model can support heterogeneous business targets.

\subsection{Ablation Study (RQ2)}
\label{sec:rq2_ablation}

To answer \textbf{RQ2}, we ablate OneModel from three levels: training strategy, input sequence, and model components, all reported on Click AUC for clarity.

\begin{figure}[t]
\centering
\includegraphics[width=0.99\columnwidth]{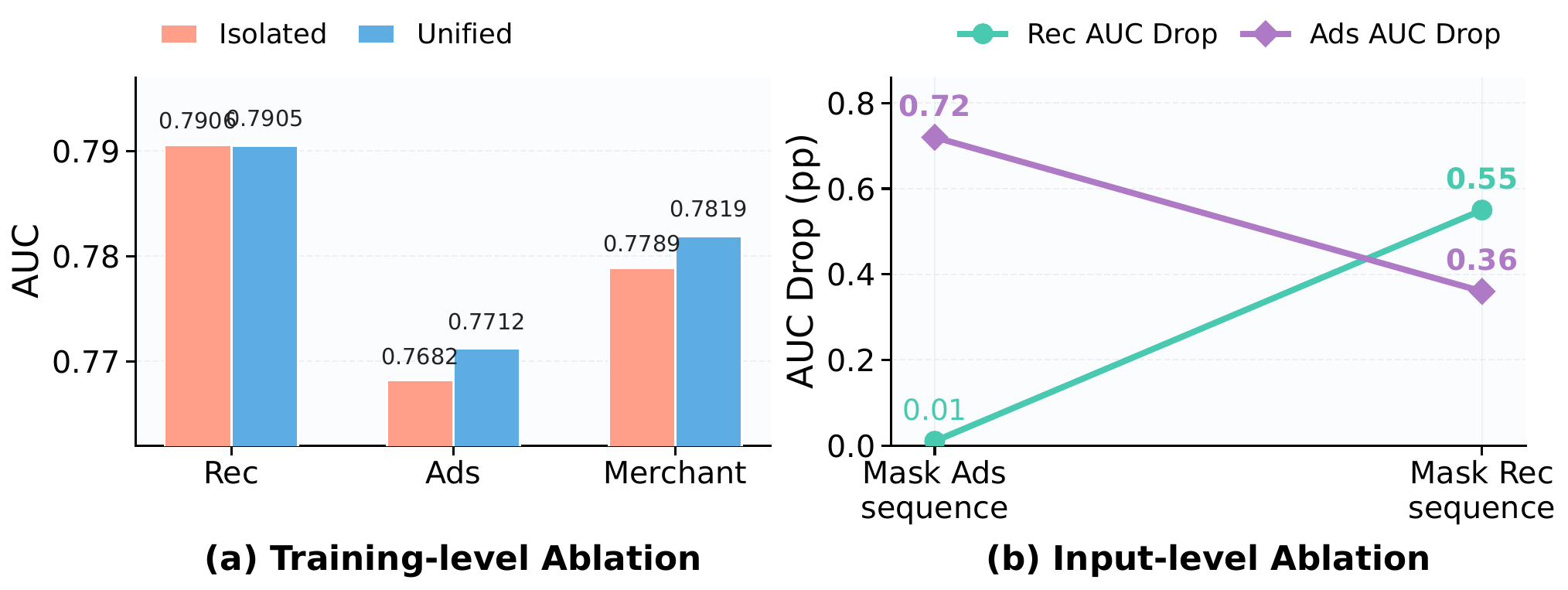}
\caption{
Training- and input-level ablations.
(a) Per-stream isolated training vs. unified multi-stream training on Rec, Ads, and Merchant Click AUC.
(b) Sub-sequence masking under unified training, reported as AUC drop on Rec and Ads targets.
}
\label{fig:rq2_training_input_ablation}
\end{figure}

\textbf{Training-level ablation.}
Figure~\ref{fig:rq2_training_input_ablation}(a) compares per-stream isolated training with unified multi-stream training.
Unified training improves Ads Click AUC from 0.7682 to 0.7712 and Merchant from 0.7789 to 0.7819, both by +3.0\textperthousand, while Rec remains nearly unchanged at 0.7905 compared with the isolated 0.7906.
This indicates that the two lower-resource streams benefit from the shared backbone, while the dominant Rec stream does not suffer meaningful negative transfer.

\textbf{Input-level ablation.}
Figure~\ref{fig:rq2_training_input_ablation}(b) masks different sub-sequences at inference time to test whether cross-scenario histories provide useful evidence.
For Ads prediction, masking its own Ads history causes the largest drop, while masking the Rec sub-sequence also reduces Ads AUC, showing that Rec behaviors provide auxiliary intent signals for Ads ranking.
In contrast, Rec prediction is only slightly affected by removing Ads history, which is consistent with Rec being the highest-resource stream.

\textbf{Module-level ablation.}
Table~\ref{tab:core_ablation} ablates three core modules under unified training on Ads Click AUC.
Removing unified representation, SAIM, and stratified user representation leads to consistent drops of 1.4\textperthousand, 1.1\textperthousand, and 0.8\textperthousand, respectively.
These results show that representation alignment, scenario-aware modulation, and multi-timescale user modeling each contribute to the final performance.

\begin{table}[t]
\centering
\small
\setlength{\tabcolsep}{6pt}
\renewcommand{\arraystretch}{1.12}
\caption{Module-level ablation study of OneModel on the Ads stream under unified multi-stream training. $\Delta$ denotes the absolute Click AUC decrease in per-mille (\textperthousand) relative to the full OneModel.}
\label{tab:core_ablation}
\begin{tabularx}{\linewidth}{Xcc}
\toprule
\rowcolor{lightgrayrow}
\textbf{Variant} & \textbf{Click AUC} & \textbf{$\Delta$ (\textperthousand)} \\
\midrule
\rowcolor{oursblue}
Full OneModel
& \textbf{0.7712} & -- \\
\midrule
w/o Unified Rep.
& 0.7698 & 1.4 \\
w/o SAIM
& 0.7701 & 1.1 \\
w/o Stratified Rep.
& 0.7704 & 0.8 \\
\bottomrule
\end{tabularx}
\end{table}

\subsection{Scaling Analysis (RQ3)}
\label{sec:rq3_scaling}

\begin{figure*}[t]
\centering
\includegraphics[width=0.98\textwidth]{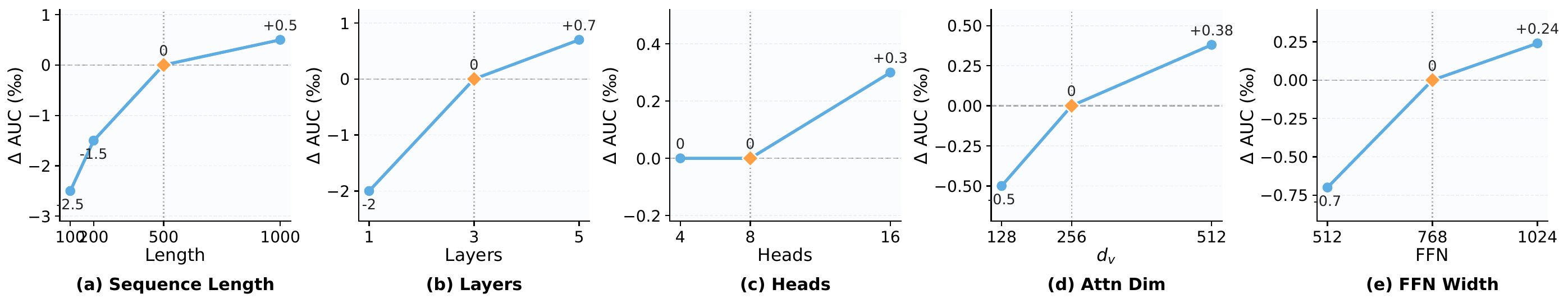}
\caption{
Scaling analysis of OneModel.
(a) Effect of input sequence length on offline AUC, reported relative to the base length 500. The diamond marker denotes the deployed configuration.
(b)--(e) Architectural sensitivity on the advertising stream, where one component is varied at a time while the others are fixed to the base configuration.
The $y$-axis reports the relative Click AUC change in per-mille (\textperthousand).
}
\label{fig:rq3_scaling_summary}
\end{figure*}

To answer \textbf{RQ3}, we analyze how OneModel scales with input sequence length and architectural capacity. Figure~\ref{fig:rq3_scaling_summary} summarizes the results: subfigure~(a) reports sequence-length scaling with length 500 as the base configuration, while subfigures~(b)--(e) vary one architectural component at a time, with the others fixed to the base configuration of 3 layers, 8 heads, $d_v{=}256$, and FFN width 768.

\textbf{Sequence length.}
Figure~\ref{fig:rq3_scaling_summary}(a) shows that shorter histories hurt performance: reducing the sequence length to 100 and 200 decreases AUC by \textbf{2.5\textperthousand} and \textbf{1.5\textperthousand}, respectively. Extending the length from 500 to 1000 brings only a small \textbf{+0.5\textperthousand} gain. We therefore use length 500 as the deployed configuration to balance accuracy and serving cost.

\textbf{Model depth.}
Figure~\ref{fig:rq3_scaling_summary}(b) shows that depth is the most sensitive dimension. Reducing the model from 3 to 1 layer drops Click AUC by \textbf{-2.0\textperthousand}, while extending it to 5 layers brings a \textbf{+0.7\textperthousand} gain.

\textbf{Attention heads.}
Figure~\ref{fig:rq3_scaling_summary}(c) shows that head count has the weakest effect. Reducing from 8 to 4 heads barely changes Click AUC, and growing to 16 heads gives only a small \textbf{+0.3\textperthousand} gain.

\textbf{Attention dimension.}
Figure~\ref{fig:rq3_scaling_summary}(d) shows that enlarging $d_v$ brings moderate improvement. Reducing $d_v$ to 128 drops AUC by \textbf{-0.5\textperthousand}, while enlarging it to 512 improves AUC by \textbf{+0.38\textperthousand}.

\textbf{FFN width.}
Figure~\ref{fig:rq3_scaling_summary}(e) shows a similar but weaker trend for FFN width. Reducing the width to 512 drops AUC by \textbf{-0.7\textperthousand}, while enlarging it to 1024 brings a \textbf{+0.24\textperthousand} gain.

\subsection{Training Stability Analysis (RQ4)}
\label{sec:rq4_stability}

To answer \textbf{RQ4}, we examine whether OneModel can be stably optimized under heterogeneous multi-stream objectives. We focus on two aspects: whether gradient isolation stabilizes shared-backbone training, and whether SAIM learns stream-dependent modulation rather than collapsing to a uniform transformation.

\begin{figure}[t]
\centering
\includegraphics[width=0.85\linewidth]{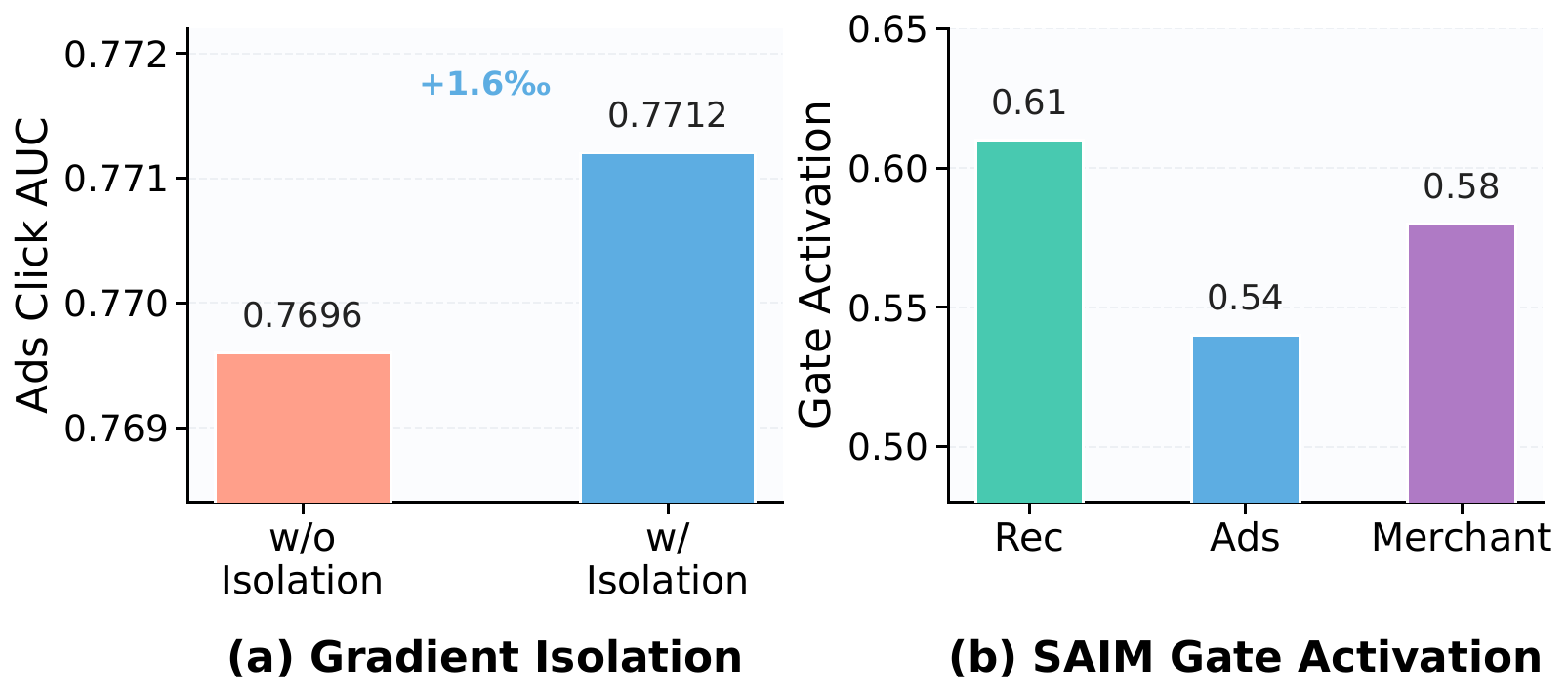}
\caption{
Training stability analysis of OneModel.
(a) Gradient isolation improves Ads Click AUC under heterogeneous multi-stream objectives.
(b) SAIM learns different average gate activations across Rec, Ads, and Merchant streams.
}
\label{fig:rq4_training_stability}
\end{figure}

\textbf{Effect of gradient isolation.}
Figure~\ref{fig:rq4_training_stability}(a) shows that removing gradient isolation decreases Ads Click AUC from 0.7712 to 0.7696.
This confirms that optimization stabilization is useful when multiple stream-specific losses update a shared backbone, since early-stage task towers may produce noisy or conflicting gradients.
Gradient isolation allows task towers to warm up before fully affecting the shared representation.

\textbf{Analysis of SAIM gate behavior.}
Figure~\ref{fig:rq4_training_stability}(b) reports the average SAIM gate activation across business streams.
The activations differ across Rec, Ads, and Merchant at 0.61, 0.54, and 0.58, respectively.
This shows that SAIM does not apply the same transformation to all streams, but learns stream-dependent channel modulation for balancing shared modeling and stream-specific specialization.

\subsection{Online Effectiveness Evaluation (RQ5)}
\label{sec:rq5_online}

To answer \textbf{RQ5}, we conduct online A/B tests against the current production ranking system across Explore Feed, Feed Advertising, and Merchant Recommendation, with users randomly assigned to control and treatment groups. As an offline sanity check, OneModel improves AUC over the production ranker by \textbf{+0.3\%} on Click and \textbf{+0.04\%} on Like, which is consistent with the online results. As summarized in Table~\ref{tab:online_ab_summary}, OneModel improves user engagement in Explore Feed, advertising value and CTR in Feed Advertising, and merchant-side conversion efficiency in Merchant Recommendation. In Merchant Recommendation, exposure PV slightly decreases by $-0.9513\%$, while DGMV, DAB, GPM, and OPM improve by \textbf{+1.1867\%}, \textbf{+1.8009\%}, \textbf{+2.1585\%}, and \textbf{+2.8118\%}, respectively, indicating higher business efficiency under fewer exposures.

\begin{table}[t]
\centering
\footnotesize
\setlength{\tabcolsep}{4pt}
\renewcommand{\arraystretch}{1.12}
\caption{Online A/B test results. Metrics are relative changes over the control group.}
\label{tab:online_ab_summary}
\begin{tabularx}{\linewidth}{lXc}
\toprule
\rowcolor{lightgrayrow}
\textbf{Metric} & \textbf{Description} & \textbf{Rel. Change} \\
\midrule
\rowcolor{gray!8}
\multicolumn{3}{l}{\textit{Explore Feed}} \\
Time Spent & User consumption time & \textbf{+0.33\%} \\
Reads & Content reading behavior & \textbf{+0.63\%} \\
Engagement & User interaction & \textbf{+1.25\%} \\
LT7 & Long-term user value & \textbf{+0.15\%} \\
\midrule
\rowcolor{gray!8}
\multicolumn{3}{l}{\textit{Feed Advertising}} \\
ADVV & Advertising value & \textbf{+3.43\%} \\
CTR & Click-through rate & \textbf{+8.18\%} \\
Impressions & Exposure volume & $-1.07\%$ \\
CPM & Cost per mille impressions & \textbf{+0.90\%} \\
\midrule
\rowcolor{gray!8}
\multicolumn{3}{l}{\textit{Merchant Recommendation}} \\
PV & Exposure volume & $-0.9513\%$ \\
DGMV & Direct GMV & \textbf{+1.1867\%} \\
DAB & Exposure-to-purchase rate & \textbf{+1.8009\%} \\
GPM & GMV per mille impressions & \textbf{+2.1585\%} \\
OPM & Purchases per mille impressions & \textbf{+2.8118\%} \\
\bottomrule
\end{tabularx}
\end{table}

\subsection{Online Cost-effectiveness Evaluation (RQ6)}
\label{sec:rq6_cost}

To answer \textbf{RQ6}, we evaluate whether OneModel can scale model capacity while remaining practical for online serving. As shown in Table~\ref{tab:online_efficiency}, OneModel increases dense parameters from 173M to 230M, but the optimized serving path reduces online latency from 270ms to 90ms. This improvement mainly comes from four system-level optimizations: feature decomposition, user feature prefetching, shared user-tower computation, and graph-level inference optimization. Feature decomposition separates reusable user-side computation from request-time candidate scoring; user feature prefetching moves part of the feature retrieval cost before the ranking request; shared user-tower computation avoids repeatedly encoding the same user state for different candidates; and graph optimization reduces redundant operators in online inference. These results show that OneModel improves model capacity while maintaining efficient production serving.

\begin{table}[t]
\centering
\small
\setlength{\tabcolsep}{4pt}
\renewcommand{\arraystretch}{1.12}
\caption{
Online cost-effectiveness comparison between the production baseline and OneModel.
}
\label{tab:online_efficiency}
\begin{tabular*}{\linewidth}{@{\extracolsep{\fill}}lccc@{}}
\toprule
\rowcolor{lightgrayrow}
\textbf{Metric} & \textbf{Baseline} & \textbf{OneModel} & \textbf{Rel. Change} \\
\midrule
Dense Params & 173M & \textbf{230M} & +32.9\% \\
Latency & 270ms & \textbf{90ms} & \textbf{-66.7\%} \\
\bottomrule
\end{tabular*}
\end{table}




\section{Conclusion}

We presented \textbf{OneModel}, a unified generative framework for multi-stream ranking that modeled interleaved user behaviors across organic recommendation, advertising, and merchant services with a shared backbone. OneModel integrated cross-stream representation learning, an action-oriented long-context encoder with SAIM, stratified user representations, and a hybrid training and decoupled serving design for efficient deployment. Offline and online A/B tests at \textit{Xiaohongshu} showed consistent gains over strong baselines and favorable scaling with model size and context length, which validated unified modeling as a practical foundation for industrial multi-stream recommendation.

\clearpage

\section*{Generative AI Use Disclosure}
We used generative AI tools only for language polishing and grammar refinement.

\bibliographystyle{ACM-Reference-Format}
\bibliography{Refs}

\appendix

\section{Additional Implementation Details}
\label{app:implementation_details}

For each business stream, we use user-side features such as demographics, long-term interest profiles, and activity statistics; item-side features such as content metadata and multi-modal representations where available; context features such as time, device, entry point, and page context; and interaction features such as clicks, likes, views, dwell time, and conversions. We adopt a timestamp-based split to avoid future information leakage, training on earlier interactions and validating and testing on later ones. We optimize the model with sampled softmax loss using in-batch negatives, temperature $0.05$, learning rate $1\mathrm{e}{-4}$, and zero weight decay. Under unified training, batches from Rec, Ads, and Merchant share the same backbone while retaining stream-specific prediction targets. For online inference, feature decomposition separates reusable user-side computation from request-time candidate scoring; user feature prefetching moves part of the feature retrieval cost before ranking; shared user-tower computation avoids repeated user encoding across candidates; and graph-level optimization removes redundant inference operators.

\section{Baseline} 
\label{Baseline}
We compare OneModel with representative long-sequence ranking backbones, including HSTU and GenRank~\cite{hstu,genrank}. All baselines are evaluated under both single-scenario and cross-scenario settings for fair comparison.

\end{document}